\documentclass[10pt, conference]{IEEEtran}
\IEEEoverridecommandlockouts
\usepackage[dvips]{graphicx}
\usepackage{subfigure}
\usepackage{epsfig, psfrag, amsmath, amssymb, amsfonts, latexsym, balance, hyperref,indentfirst,bookmark}
\usepackage[numbers,sort&compress]{natbib}

\begin{document}

\title{A Stochastic Geometry Approach to Energy Efficiency in Relay-Assisted Cellular Networks}
\author{Na Deng, Sihai Zhang, Wuyang Zhou, and Jinkang Zhu \\
Dept. of Elect. Eng. \& Inform. Sci., University of Science \& Technology of China, China, 230027. \\
Email: {ndeng@mail.ustc.edu.cn, \{shzhang, wyzhou, jkzhu\}@ustc.edu.cn}.
\thanks{This work was partially supported by NSFC (61172088), National programs for High Technology Research and Development (SS2012AA011702), the National Major Special Projects in Science and Technology of China under grant 2010ZX03003-001 and 2010ZX03005-003.
}}

\maketitle
\thispagestyle{empty}
\begin{abstract}
Though cooperative relaying is believed to be a promising technology to improve the energy efficiency of cellular networks, the relays' static power consumption might worsen the energy efficiency therefore can not be neglected. In this paper, we focus on whether and how the energy efficiency of cellular networks can be improved via relays. Based on the spatial Poisson point process, an analytical model is proposed to evaluate the energy efficiency of relay-assisted cellular networks. With the aid of the technical tools of stochastic geometry, we derive the distributions of signal-to-interference-plus-noise ratios (SINRs) and mean achievable rates of both non-cooperative users and cooperative users. The energy efficiency measured by ``bps/Hz/W'' is expressed subsequently. These established expressions are amenable to numerical evaluation and corroborated by simulation results.
\end{abstract}

\setcounter{page}{1}

\section{Introduction}
\label{sec:intro}
With the rapid proliferation of user equipments (UEs) and base stations (BSs), the energy consumption of cellular networks grows quite amazing. It is shown in \cite{marsan2009optimal} that the energy consumed by BSs accounts for nearly 60-70\% of the total network energy. Therefore, improving the energy efficiency of BSs becomes the fundamental challenge to realize green communications.

Recently, cooperative relaying has been in-depth studied as a promising way to reduce the energy consumption of cellular networks \cite{fantini2011energy}\cite{lee2010minimum}, which can extend coverage, increase capacity and provide flexible cost-effective deployment options as well \cite{3gpp2009further}. Compared with BSs, RSs cover a much smaller area and require lower transmit power. Meanwhile, the UEs covered by RSs mostly enjoy much higher average signal-to-interference-plus-noise ratios (SINRs). As RSs do not have a wired backhaul connection, their deployment cost is largely reduced. Consequently, using cooperative relaying technique is capable of lowering the transmit power consumption in cellular networks and the low power RSs are easy to be deployed without modifying the current cellular infrastructure.

Generally, the total power consumption of both BSs and RSs consists of two parts, i.e., the transmit power consumption and the static power consumption. Compared with the transmit power consumption, the static power consumption which is contributed by signal processing, battery backup, site cooling and etc., occupies a substantial part of the whole power consumption. Therefore, the transmit power saved by cooperative relaying technique may not compensate for the additional power consumption of RSs which leads to higher power consumption in cellular networks.

The aforementioned discussions motivate us to investigate whether and how the energy efficiency of cellular networks can be improved via relays. The related works mostly focus on regular network deployments, such as the hexagonal grid model, which are merely estimated by Monte Carlo simulations \cite{fantini2011energy}\cite{lei2010opportunistic}\cite{fehske2009energy}.
However, the spatial deployment observed in actual communication networks is usually far from being regular. In addition, the interference and the signal strength at a receiver critically depend on the spatial positions of the interfering transmitters, thereby mathematical techniques are needed to explicitly model the node distribution. Stochastic geometry theory has recently emerged as an essential tool to model and quantify interference, mean achievable rate, and coverage in cellular networks which are verified to be close to the actual networks \cite{andrews2010tractable}. Consequently, this paper proposes an analytical model to evaluate the energy efficiency of relay-assisted cellular networks using stochastic geometry, instead of relying on system simulations only. The distributions of SINRs and mean achievable rates of both non-cooperative UEs and cooperative UEs are derived. We use the ratio of spectral efficiency to power consumption with unit bps/Hz/W as the metric of energy efficiency and further present the theoretical expression. Simulation results present how the intensity of users affects the energy efficiency and how much RS's static power consumption should be controlled below in order to obtain some energy saving gains. Notably, our work serves as a pioneering effort on the network design and planning, especially with respect to energy efficient wireless communications.

\section{Network Model and Metrics}
\label{sect:Network Model and Metrics}
\begin{figure}[t]
\centering
\includegraphics[width=2.9in]{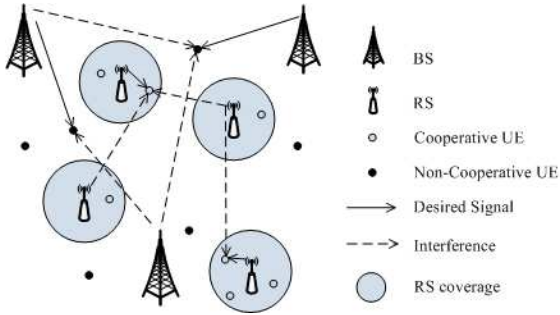}
\caption{Downlink relay-assisted cellular network model.}
\label{fig:scenario1}
\end{figure}

\begin{figure}[t]
\centering
\includegraphics[width=2.9in]{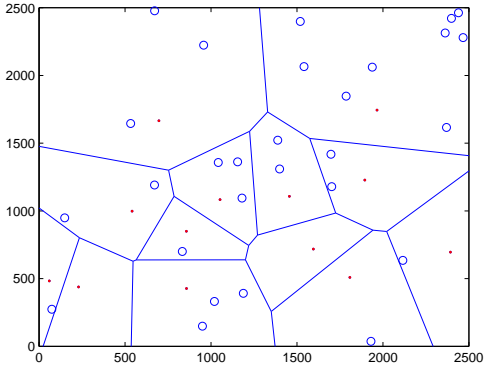}
\caption{A Voronoi network topology, where each Voronoi cell is the coverage area of a BS and each small circle represents a relay coverage area.}
\label{fig:scenario2}
\end{figure}

We consider a relay-assisted downlink cellular network shown in Fig. \ref{fig:scenario1}. The locations of BSs are modeled according to a homogeneous Poisson point
process (PPP) $\Phi_{b}$ with intensity $\lambda_{b}$, and RSs are located according to another homogeneous PPP $\Phi_{r}$ with intensity $\lambda_{r}$. Specifically, for a given PPP, the number of points in a bounded area is a Poisson-distributed random variable and the points are uniformly-distributed within the area.
A realization of the relay-assisted cellular network with Poisson distributed BSs and relays portrayed by a Voronoi tessellation is given in Fig. \ref{fig:scenario2}. Here, we use the term \emph{cell} to refer to a Voronoi region with a random area \cite{haenggi2009stochastic} of each BS, which, as noted in \cite{brown2000cellular}, more closely reflects actual deployments that are highly non-regular.
We denote the area of the Voronoi cell by $S$. There is no known closed-form expression of the probability density function (PDF) of $S$, whereas a simple approximation \cite{ferenc2007size} has been proved sufficiently accurate for practical purposes. Considering scaling, the approximate PDF of the size of a single-cell coverage area is given by
\begin{equation} \label{1}
f(S)=\frac{343}{15}\sqrt{\frac{7}{2\pi}}(S\lambda_{b})^\frac{5}{2}\exp(-\frac{7}{2}S\lambda_{b})\lambda_{b}
\end{equation}
Each RS has a circular coverage with radius $R_{r}$. The cooperative UEs are distributed according to a homogeneous PPP with intensity $\lambda_{c}$ in each circular area, and the non-cooperative UEs are distributed according to another homogeneous PPP with intensity $\lambda_{nc}$ in the whole network. All the PPPs are mutually independent.

We assume that each UE is served by its closest base station which is at a distance $r_{B}$. Since the BSs are distributed as PPPs, it follows that $r_{B}$ is Rayleigh
distributed with parameter $1/\sqrt{2\pi\lambda_{b}}$ \cite{stoyan1995stochastic}. If the UE is also located in a RS's coverage, the distance $r_{R}$ between the RS and
the UE follows the distribution $f_{R}(r)=2r/R_{r}^{2}$. The transmit power is $P_{B}$ for all the BSs and $P_{R}$ for all the RSs. For the sake of convenience, we adopt a standard path loss propagation model with path loss exponent $\alpha>2$. Regarding fading, we assume that the link between the serving access point (either a BS or a RS) and the served UE experiences Rayleigh fading with parameter $\mu$.
The set of interfering entities is the ones that use the same subchannel as the served UE and the fading of interference links may follow an arbitrary probability distribution, which is denoted by $g$. For all receivers, the noise power is $\sigma^{2}$.

The different PPPs corresponding to different entities in the network interact in a complicated way, which makes a rigorous statistical analysis extremely difficult. For example, a cooperative UE may be covered by more than one RS that leads to the delicate issue of relay selection, and furthermore renders the distributions of subchannel usage among RSs and BSs intrinsically correlated. To overcome the technical difficulties due to spatial interactions, we propose a two-scale approximation of the network model for the subsequent analysis motivated by the fact that the covered area of a RS is significantly smaller than that of a BS. The two-scale approximation consists of two views, i.e., the macro-scale view and the micro-scale view. The macro-scale view concerns an observer outside the coverage area of a RS, and the whole coverage area of the RS shrinks to a single point, marked by the number of cooperative UEs. The micro-scale view concerns an observer inside the coverage area of a RS, and the coverage area is still a circular with radius $R_{r}$ where the cooperative UEs are spatially distributed. By such a two-scale approximation, a cooperative UE can only be covered by a unique RS, and the coverage area of a RS can only be within a unique Voronoi cell of a BS.

\subsection{Distribution of UEs}
\label{subsec:Distribution of UEs}
Let $U_{c}$ be the number of cooperative UEs in a RS's coverage and from our model we have $U_{c}\sim Poisson(\lambda_{c}\pi R_{r}^{2})$. The number of non-cooperative users served only by BSs is denoted by $U_{nc}$, and it is characterized as follows. Let $S$ be an arbitrary single-cell coverage area, then the number of non-cooperative users in this cell is a Poisson random variable with mean $\lambda_{nc}S$. Conditioned upon $S$, the probability generating function (PGF) of the unconditioned $U_{nc}$ is thus given by
\begin{eqnarray} \label{2}
G(z)\!\!\!\!\!&=\!\!\!\!\!&\int_{0}^{\infty}\exp(\lambda_{nc}(z-1)S)f(S)dS  \nonumber \\
    \!\!\!\!\!&=\!\!\!\!\!&\frac{343}{8}\sqrt{\frac{7}{2}}\left(\frac{7}{2}-\frac{\lambda_{nc}}{\lambda_{b}}(z-1)\right)^{-\frac{7}{2}}
\end{eqnarray}
where $f(S)$ is given by (\ref{1}), and the distribution of $U_{nc}$ is therefore given by the derivatives of $G(z)$,
\begin{equation} \label{3}
\mathrm{P}\{U_{nc}=i\}=\frac{G^{(i)}(0)}{i!}, ~~~i=0,1,...
\end{equation}
Similarly, we can obtain $\mathrm{P}\{U_{r}=i\}$, where $U_{r}$ represents the number of relays in a Voronoi cell with mean $\lambda_{r}S$.

\subsection{Spectrum Allocation and Usage}
\label{subsec:Spectrum Allocation and Usage}
The available spectrum for a cell is evenly divided into $M$ subchannels. We assume that RSs are given a separate frequency band from BSs, such that there is no cross-tier interference. Thus a partition $(M_{b},M_{r})$ is made, where $M_{b}$ subchannels are available for BS transmissions and $M_{r}=M-M_{b}$ subchannels are available for RS transmissions. When the current UEs are more than the number of subchannels in a serving access point, they are served by time-sharing with equal time proportion. To distinguish between non-cooperative UEs and cooperative UEs for BS, $M_{b}$ are further divided into two parts, i.e., $M_{b_{1}}$ and $M_{b_{2}}$, which are used for non-cooperative transmissions and cooperative transmissions, respectively. From the macro-scale view, for a BS, the access required from the cooperative UEs is equivalent to that from the RS. In this case, we divide $M_{b_{2}}$ into $\rho$ parts, where $\rho=M_{b_{2}}/M_{r}$, and due to the broadcast nature of wireless medium, each part of the subchannels are used for both BS-to-RS link and BS-to-UE links.

The subchannels in each partition ($M_{b_{1}}$, $M_{b_{2}}$, or $M_{r}$) are uniformly and independently selected by BS or RSs, and it suffices to analyze an arbitrary one of them. Firstly, let us examine the probability that a given subchannel in $M_{b_{1}}$ is used by a BS, which can be obtained via normalizing the average number of subchannels used by the total number of subchannels $M_{b_{1}}$. The same goes for $M_{b_{2}}$ and $M_{r}$ given as follows.
\begin{eqnarray}
P_{busy,b_{1}}=\frac{1}{M_{b_{1}}}\sum\limits_{i=0}^{\infty}\min(i,M_{b_{1}})\mathrm{P}\{U_{nc}=i\} \\
P_{busy,b_{2}}=\frac{1}{\rho}\sum\limits_{i=0}^{\infty}\min(i,\rho)\mathrm{P}\{U_{r}=i\} \\
P_{busy,r}=\frac{1}{M_{r}}\sum\limits_{i=0}^{\infty}\min(i,M_{r})\mathrm{P}\{U_{c}=i\}
\end{eqnarray}

The spatial process of RSs that use a given subchannel is the independent thinning of the original PPP of RSs $\Phi_{r}$ by the probability $\!P_{busy,r}$, denoted by $\!\Phi_{r}^{'}\!$ with intensity $\lambda_{r}^{'}\!=\!\lambda_{r}P_{busy,r}$. The term ``independent thinning'' means that $\Phi_{r}^{'}\!$ can be viewed as obtained from $\Phi_{r}\!$ by independently removing points with probability $\!1\!-\!P_{busy,r}$. Similarly, the spatial process of BSs that use a given subchannel (either in $M_{b_{1}}$ or $M_{b_{2}}$) is the independent thinning of the original PPP of BSs $\Phi_{b}$ by the probability $P_{busy,b_{1}}\!$ (or $P_{busy,b_{2}}$), denoted by $\Phi_{b}^{'}\!$ with intensity $\lambda_{b}^{'}\!=\!\lambda_{b}P_{busy,b_{1}}$ (or $\lambda_{b}^{'}\!=\!\lambda_{b}P_{busy,b_{2}}$). These two independently thinned PPPs will prove useful in the subsequent analysis.

\subsection{Power Model}
\label{subsec:Power Model}
We model the relation between the average power consumption and the average radiated power per site in a linear fashion by
\begin{equation}
\widetilde{P}_{i}=a_{i}P_{i}+b_{i}, ~for~i=B,R
\end{equation}
where $\widetilde{P}_{i}$ and $P_{i}$ denote the average consumed power and the radiated power per site, respectively, while $B$, $R$ represent BS and relay. The coefficients $a_{i}$ accounts for the power consumption that scales with the average radiated power due to amplifier and feeder losses as well as cooling of sites. The terms $b_{i}$ denotes power offsets which are consumed independently of the average transmit power. These offsets are, amongst others, due to signal processing, battery backup, as well as site cooling \cite{fehske2009energy}.

\subsection{Energy-Efficiency Function}
\label{subsec:Energy Efficiency Function}
The energy-efficiency evaluation function of cellular networks in this paper is defined as the ratio of spectral efficiency to power consumption with unit bps/Hz/W. Here we use the mean achievable rate to measure the spectral efficiency. According to our model, the cellular network is homogeneous and the energy efficiency of the whole network is equivalent to the energy efficiency of an arbitrary Voronoi cell which can be expressed as
\begin{equation} \label{8}
Q=\frac{\tau}{\widetilde{P}_{B}+N\widetilde{P}_{R}}
\end{equation}
where $\tau$ denotes the cell's total mean achievable rate provided by both BS with consumed power $\widetilde{P}_{B}$ and RSs with consumed power $N\!\widetilde{P}_{R}$, and $N$ is the average number of RSs in the cell.

\section{Analysis of Mean Achievable Rate and Energy Efficiency}
\label{Analysis of SINR and Mean Achievable Rate}
In this section, we derive the mean achievable rates of cooperative UEs and non-cooperative UEs at first, and then give the energy efficiency expression based on the definition in \ref{subsec:Energy Efficiency Function}. For each type of UEs, we begin with general settings, and then simplify the general results under several specific parameters. The mean achievable rates are the averaged instantaneous achievable rates over both channel fading and spatial distributions of UEs and access points.

\subsection{Non-Cooperative UEs}
\label{sect:Non-Cooperative UEs}

\subsubsection{General Case}
\label{sect:General Case}
The UE is served by occupying one of subchannels with transmit power $P_{b}\!=\!\frac{P_{B}}{M_{b}}$ in the BS at a distance $r$, $g.$ is the fading of an interference link, and $R.$ is the distance between the UE and an interfering access point, thus the SINR experienced by the UE is given by $\!\gamma\!=\!\frac{P_{b}hr^{-\alpha}}{I_{b}+\sigma^{2}}$, where $h\!\sim\!exp(\mu)$, $I_{b}\!=\!\Sigma_{i\in\Phi_{b}^{'}\setminus{b_{0}}}P_{b}g_{i}R_{i}^{-\alpha}$ is the interference from the ``BS'' tier (excluding the serving BS itself which is denoted by $b_{0}$). We can thus derive the CDF of $\gamma$ as
\begin{eqnarray} \label{9}
&&Pr(\gamma\leq T)=1-Pr(\gamma>T) \nonumber \\
&=&1-\int_{0}^{\infty}Pr\left(h>\frac{Tr^{\alpha}(I_{b}+\sigma^{2})}{P_{b}}\right)f(r)dr \nonumber \\
&=&1-\int_{0}^{\infty}e^{-\frac{\mu Tr^{\alpha}\sigma^{2}}{P_{b}}}\mathcal{L}_{I_{b}}\left(\frac{\mu Tr^{\alpha}}{P_{b}}\right)f(r)dr
\end{eqnarray}
where $f(r)=2\pi\lambda_{b}re^{-\lambda_{b}\pi r^{2}}$.
By the definition of Laplace transform we get
\begin{eqnarray} \label{10}
&&\mathcal{L}_{I_{b}}(s)=\mathbb{E}_{I_{b}}\left[e^{-sI_{b}}\right] \nonumber \\
&=&\mathbb{E}_{\Phi_{b}^{'},{g_{i}}}\Bigg[\exp\Bigg(-s\!\!\!\!\!\!\sum\limits_{i\in\Phi_{b}^{'}\setminus\{b_{0}\}}\!\!\!\!\!\!P_{b}g_{i}R_{i}^{-\alpha}\Bigg)\Bigg] \nonumber \\
&=&\exp(-2\pi\lambda_{b}^{'}\underbrace{\int_{r}^{\infty}(1-\mathcal{L}_{g}(sv^{-\alpha}P_{b}))vdv}_{(A)})
\end{eqnarray}
where the last equality follows from the PGF of PPP \cite{stoyan1995stochastic}. By exchanging the integral order, we obtain
\begin{equation} \label{11}
\!\!\!\!\!(A)\!=\!-\frac{r^{2}}{2}\!+\!\frac{(sP_{b})^\frac{2}{\alpha}}{\alpha}\mathbb{E}_{g}\!\Big[g^{\frac{2}{\alpha}}\!\Big(\Gamma(-\frac{2}{\alpha},sP_{b}gr^{-\alpha})\!-\!\Gamma(-\frac{2}{\alpha})\Big)\Big]
\end{equation}
Then we can obtain the CDF of $\gamma$ by substituting (\ref{10}) and (\ref{11}) into (\ref{9}).

Now we evaluate the mean achievable rate. Since for a positive random variable $X$, $\mathbb{E}[X]=\int_{t>0}\mathrm{P}\{X>t\}dt$, we have
\begin{eqnarray} \label{12}
&&\tau_{nc}=\mathbb{E}[\ln(1+\gamma)] \nonumber \\
&=&\!\!\!\!\!\int_{0}^{\infty}\!\!\!\int_{0}^{\infty}Pr\left(h>\frac{(I_{b}+\sigma^{2})(e^{t}-1)}{P_{b}r^{-\alpha}}\right)dtf(r)dr \nonumber \\
&=&\!\!\!\!\!\int_{0}^{\infty}\!\!\!\int_{0}^{\infty}e^{-\frac{\mu r^{\alpha}\sigma^{2}(e^{t}-1)}{P_{b}}}\mathcal{L}_{I_{b}}\left(\frac{\mu r^{\alpha}(e^{t}-1)}{P_{b}}\right)dtf(r)dr \nonumber \\
&=&\!\!\!\!\!\pi\lambda_{b}\!\!\!\int_{0}^{\infty}\!\!\!\int_{0}^{\infty}\exp\left(-\frac{\mu v^{\frac{\alpha}{2}}\sigma^{2}(e^{t}-1)}{P_{b}}+ \right. \nonumber \\
&&\left. \pi\lambda_{b}^{'}v\Big(1-\varphi(e^{t}-1,\alpha)-\frac{1}{P_{busy,b_{1}}}\Big)\right)dtdv
\end{eqnarray}
where $\varphi(T,\alpha)$ is given by
\begin{equation}
\!\!\!\varphi(T,\alpha)=\frac{2}{\alpha}(\mu T)^{\frac{2}{\alpha}}\mathbb{E}_{g}\left[g^{\frac{2}{\alpha}}\Big(\Gamma(-\frac{2}{\alpha},\mu Tg)-\Gamma(-\frac{2}{\alpha})\Big)\right]
\end{equation}
The integrals in (\ref{12}) can be evaluated by numerical methods, while they can be simplified to a concise form in the special case.

\subsubsection{Special Case}
\label{sect:Special Case}
When the interference experiences Rayleigh fading with mean $\mu$, i.e., $g\sim exp(\mu)$, and $\alpha=4$. The results are simplified as follows.
\begin{eqnarray}
&&\!\!\!\!\!Pr(\gamma\!\leq \!T)= \nonumber \\
&&1-\pi\lambda_{b}\sqrt{\frac{\pi P_{b}}{\mu T\sigma^{2}}}\exp\left(\frac{\kappa^{2}(T)P_{b}}{4\mu T\sigma^{2}}\right)Q\left(\kappa(T)\sqrt{\frac{P_{b}}{2\mu T\sigma^{2}}}\right) \nonumber \\
\end{eqnarray}
\begin{eqnarray}
\tau_{nc}=\pi\lambda_{b}\int_{0}^{\infty}\sqrt{\frac{\pi P_{b}}{\mu \sigma^{2}(e^{t}-1)}}\exp\left(\frac{\kappa^{2}(e^{t}-1)P_{b}}{4\mu \sigma^{2}(e^{t}-1)}\right)\times \nonumber \\
Q\left(\kappa(e^{t}-1)\sqrt{\frac{P_{b}}{2\mu \sigma^{2}(e^{t}-1)}}\right)dt
\end{eqnarray}
where $Q(x)=\frac{1}{\sqrt{2\pi}}\int_{x}^{\infty}e^{-y^{2}/2}dy$, and $\kappa(T)$ is given by
\begin{equation}
\kappa(T)=\pi\lambda_{b}\left(1+P_{busy,b_{1}}\sqrt{T}\Big(\frac{\pi}{2}-\arctan\frac{1}{\sqrt{T}}\Big)\right)
\end{equation}

\subsection{Cooperative UEs}
\label{sect:Coded Cooperation Transmission}

\subsubsection{General Case}
\label{sect:General Case}
To obtain the mean achievable rate of a cooperative UE, we have to derive the distributions of SINRs ($\gamma_{0}$, $\gamma_{1}$, $\gamma_{2}$) and mean achievable rates ($\tau_{0}$, $\tau_{1}$, $\tau_{2}$) for BS-to-UE link, BS-to-RS link and RS-to-UE link, respectively.
Based on our two-scale approximation, we use $\gamma_{1}$ to approximate $\gamma_{0}$ in the subsequent analysis. Fig. \ref{fig:SINR} in \ref{sect:numerical} verifies the reasonableness of this approximation. The transmission scheme termed coded cooperation with selective source-to-destination parity transmission \cite{deng2012novel} is introduced here for cooperative UEs. The scheme has two modes depending on whether the RS decodes the message from BS correctly. Accordingly, different modes have different mean achievable rates. The specific analysis is presented below.

The probability of correctly decoding for the RS is denoted by $P_{d}$ and expressed as
\begin{eqnarray}
P_{d}=Pr(\gamma_{1}\geq T_{th})=\pi\lambda_{b}\!\!\!\int_{0}^{\infty}\exp\left(-\frac{\mu v^{\frac{\alpha}{2}}\sigma^{2}T_{th}}{P_{b}} \right. \nonumber \\
\left. +\pi\lambda_{b}^{'}v\Big(1-\varphi(T_{th},\alpha)-\frac{1}{P_{busy,b_{2}}}\Big)\right)dv
\end{eqnarray}
where $T_{th}\!\!=\!\!2^{R_{th}/\beta}\!-\!1$, $\!R_{th}\!$ is the target rate of RS correctly decoding and $\beta$ is the cooperative parameter in coded cooperation.

\paragraph{Mode 1}
\label{sect:Mode 1}
When the RS fails to decode, the cooperation is degraded to a direct transmission.
\begin{equation} \label{18}
Pr(\gamma_{1}\!\leq\!T\!\mid\!\gamma_{1}\!<\!T_{th})=
\left\{ \begin{matrix}
   {1, ~~~~~~~~~~~~~T\geq T_{th}}  \\
   {\frac{Pr(\gamma_{1}\leq T)}{Pr(\gamma_{1}<T_{th})}, ~T<T_{th}}  \\
\end{matrix} \right.
\end{equation}
The conditional PDF of $\gamma_{1}$ denoted by $f_{\gamma_{1}}(t\!\!\mid\!\!\gamma_{1}\!\!<\!\!T_{th})$ is obtained by differentiating (\ref{18}) with respect to $T$. Then we get the mean achievable rate in this mode as
\begin{eqnarray} \label{19}
\tau_{m1}\!\!\!&=&\!\!\!\int_{0}^{T_{th}}\!\!\!ln(1+t)f_{\gamma_{1}}(t\!\mid\!\gamma_{1}\!<\!T_{th})dt \nonumber \\
\!\!\!&=&\!\!\!ln(1+T_{th})\!-\!\frac{1}{1-P_{d}}\int_{0}^{T_{th}}\!\!\frac{Pr(\gamma_{1}<t)}{1+t}dt
\end{eqnarray}

\paragraph{Mode 2}
\label{sect:Mode 2}
When the RS decodes correctly, the data is transmitted from BS to UE and the parity is transmitted from RS to UE.

For BS-to-RS link, the derivation of $\tau_{1}$ is similar to non-cooperative UEs in \ref{sect:Non-Cooperative UEs}, by just revising $P_{busy,b_{1}}$ to $P_{busy,b_{2}}$.

For BS-to-UE link,
\begin{equation} \label{20}
Pr(\gamma_{1}\!\leq\!T\!\mid\!\gamma_{1}\!\geq\!T_{th})=
\left\{ \begin{matrix}
   {0, ~~~~~~~~~~~~~~~~~~~~~~~~~T<T_{th}}  \\
   {\frac{Pr(\gamma_{1}\leq T)-Pr(\gamma_{1}<T_{th})}{1-Pr(\gamma_{1}<T_{th})}, ~T\geq T_{th}}  \\
\end{matrix} \right.
\end{equation}
The conditional PDF of $\gamma_{1}$ denoted by $f_{\gamma_{1}}(t\!\!\mid\!\!\gamma_{1}\!\!\geq\!\!T_{th})$ is obtained by differentiating (\ref{20}) with respect to $T$. Then we get the mean achievable rate of BS-to-UE link as follows
\begin{eqnarray}
\tau_{0}\!\!\!\!&=&\!\!\!\!\int_{T_{th}}^{\infty}\!\!\!ln(1+t)f_{\gamma_{1}}(t\!\mid\!\gamma_{1}\!\geq\!T_{th})dt \nonumber \\
\!\!\!\!&=&\!\!\!\!\frac{1}{P_{d}}(\tau_{1}-(1-P_{d})\tau_{m1})
\end{eqnarray}

For RS-to-UE link, the UE is served by occupying one of subchannels with transmit power $P_{r}\!=\!\frac{P_{R}}{M_{r}}$ in the RS at a distance $r$. Denote by $I_{r}$ the interference strengths from the ``RS'' tier.
The CDF of $\gamma_{2}$ is given by
\begin{equation} \label{22}
Pr(\gamma_{2}\leq T)=1-\int_{0}^{R_{r}}e^{-\frac{\mu Tr^{\alpha}\sigma^{2}}{P_{r}}}\mathcal{L}_{I_{r}}(\frac{\mu Tr^{\alpha}}{P_{r}})f(r)dr
\end{equation}
Following the same procedure as in the derivation of (10), we get the Laplace transform of $I_{r}$ as
\begin{equation} \label{23}
\mathcal{L}_{I_{r}}(s)=\exp\left(-\pi\lambda_{r}^{'}(sP_{r})^{\frac{2}{\alpha}}\Gamma(1-\frac{2}{\alpha})\mathbb{E}_{g}(g^{\frac{2}{\alpha}})\right)
\end{equation}
Substituting (\ref{23}) into (\ref{22}) with $v=r^{2}$, we get the distribution of $\gamma_{2}$. Similar to (\ref{12}), $\tau_{2}$ is given by
\begin{eqnarray} \label{24}
\!\!\!\!\!\!\tau_{2}\!\!\!\!&=&\!\!\!\!\!\int_{0}^{R_{r}}\!\!\!\!\!\int_{0}^{\infty}\!\!e^{-\frac{\mu r^{\alpha}\sigma^{2}(e^{t}-1)}{P_{r}}}\!\mathcal{L}_{I_{r}}\!\left(\frac{\mu r^{\alpha}(e^{t}-1)}{P_{r}}\right)\!dtf(r)dr \nonumber \\
&=&\!\!\!\!\!\frac{1}{R_{r}^{2}}\!\!\int_{0}^{R_{r}^{2}}\!\!\!\!\!\int_{0}^{\infty}\!\!\!\!\exp\!\left(\!-\frac{\mu v^{\frac{\alpha}{2}}\sigma^{2}(e^{t}-1)}{P_{r}}\!+\!\psi(\alpha)(e^{t}-1)^{\frac{2}{\alpha}}v\!\right)\!dtdv \nonumber \\
\end{eqnarray}
where $\psi(\alpha)$ is given by
\begin{equation}
\psi(\alpha)=\frac{2}{\alpha}\pi\mu^{\frac{2}{\alpha}}\lambda_{r}^{'}\Gamma(-\frac{2}{\alpha})\mathbb{E}_{g}(g^{\frac{2}{\alpha}})
\end{equation}
According to the capacity formula in \cite{deng2012novel}, we get the mean achievable rate in this mode below
\begin{equation}
\tau_{m2}=\beta\tau_{0}+(1-\beta)\tau_{2}
\end{equation}
Integrating $\tau_{m1}$ and $\tau_{m2}$, the mean achievable rate for a cooperative UE is obtained as
\begin{equation}
\tau_{c}=(1-P_{d})\tau_{m1}+P_{d}\tau_{m2}
\end{equation}

\subsubsection{Special Case}
\label{sect:Special Case}
When the interference experiences Rayleigh fading and $\alpha=4$, the results are simplified as
\begin{eqnarray}
\!\!\!\!\!\!\!\!P_{d}\!=\!\pi\!\lambda_{b}\!\sqrt{\!\frac{\pi P_{b}}{\mu T_{th}\sigma^{2}}}\!\exp\!\Big(\!\frac{\kappa^{2}\!(T_{th})P_{b}}{4\mu T_{th}\sigma^{2}}\!\Big)Q\Big(\!\kappa(T_{th})\!\sqrt{\!\frac{P_{b}}{2\mu T_{th}\sigma^{2}}}\Big)
\end{eqnarray}
\begin{eqnarray}
Pr(\gamma_{2}\!\leq \!T)= 1-\frac{1}{R_{r}^{2}}\sqrt{\frac{\pi P_{r}}{\mu T\sigma^{2}}}\exp\!\Big(\!\frac{\pi^{4}(\lambda_{r}^{'})^{2}\!P_{r}}{16\mu\sigma^{2}}\!\Big) \nonumber \\
\times\Bigg(\!Q\Big(\pi^{2}\!\lambda_{r}^{'}\sqrt{\frac{P_{r}}{8\mu\sigma^{2}}}\Big)\!-Q\Big(\pi^{2}\!\lambda_{r}^{'}\sqrt{\frac{P_{r}}{8\mu\sigma^{2}}}+\!R_{r}^{2}\sqrt{\frac{2\mu T\!\sigma^{2}}{P_{r}}}\Big)\!\Bigg)
\end{eqnarray}
\begin{eqnarray}
\!\!\!\!\!\tau_{2}=\!\frac{1}{R_{r}^{2}}\!\!\int_{0}^{\infty}\!\!\!\!\!\sqrt{\frac{\pi P_{r}}{\mu \sigma^{2}(e^{t}-1)}}\!\exp\!\Big(\frac{P_{r}\pi^{4}(\lambda_{r}^{'})^{2}}{16\mu\sigma^{2}}\Big)\Bigg(\!Q\Big(\frac{\pi^{2}\lambda_{r}^{'}}{2\sqrt{\frac{2\mu\sigma^{2}}{P_{r}}}}\Big)
\nonumber \\
-Q\Big(\frac{\pi^{2}\lambda_{r}^{'}}{2\sqrt{\frac{2\mu\sigma^{2}}{P_{r}}}}+R_{r}^{2}\sqrt{\frac{2\mu\sigma^{2}}{P_{r}}(e^{t}-1)}\Big)\!\Bigg)dt
\end{eqnarray}

\subsection{Energy Efficiency}
The energy efficiency in our network is expressed as
\begin{equation}
Q=\frac{\tau_{s1}+\tau_{s2}}{\widetilde{P}_{B}+N\widetilde{P}_{R}}
\end{equation}
where $\tau_{s1}$ and $\tau_{s2}$ are given by
\begin{equation}
\tau_{s1}=\sum\limits_{i=1}^{+\infty}\min(M_{b_{1}},i)\mathrm{P}\{U_{nc}=i\}\tau_{nc}
\end{equation}
\begin{equation}
\!\!\!\tau_{s2}\!=\!\!\sum\limits_{i=1}^{+\infty}\!\sum\limits_{j=1}^{+\infty}\!\min(\rho,i)\!\min(M_{r},j)\mathrm{P}\{U_{r}\!=i\}\mathrm{P}\{U_{c}\!=j\}\tau_{c}
\end{equation}

\section{Numerical Results}
\label{sect:numerical}
In this section, we present comparison of the Monte Carlo simulations and the analytical results to evaluate the energy efficiency of a downlink relay-assisted cellular network. The default configurations of system model are shown in Table \ref{table 1}.
\begin{table}[b]
  \caption{System Parameters} \label{table 1}
  \centering
\begin{tabular}{|l|l|l|}
\hline
Symbol & Description & Value \\
\hline
$\lambda_{b}$ & density of BSs & $10^{-5}$ $BS/m^{2}$ \\
\hline
$\lambda_{r}$ & density of RSs & $9\times10^{-5}$ $RS/m^{2}$ \\
\hline
$P_{B}/P_{R}$ & transmit power & 43dBm/33dBm (BS/RS) \\
\hline
M & number of subchannels & 300 ($M_{b}$=285, $M_{r}$=15) \\
\hline
$R_{r}$ & radius of RS & 20m \\
\hline
$\alpha$ & pathloss exponent & 4 \\
\hline
$\mu$ & Rayleigh fading parameters & 1 (normalized) \\
\hline
$\sigma^{2}$ & noise power & -80dBm \\
\hline
$\beta$ & cooperative parameter & 0.6 \\
\hline
$R_{th}$ & target rate of RS correctly decoding & 0.5bit/s/Hz \\
\hline
$a_{B},b_{B}$ & power model parameters of BSs & 22.6, 412.4W \cite{fehske2009energy} \\
\hline
$a_{R},b_{R}$ & power model parameters of RSs & 5.5, 32.0W \cite{fehske2009energy} \\
\hline
\end{tabular}
\end{table}

\begin{figure} [t]
\centering
\includegraphics[width=2.9in]{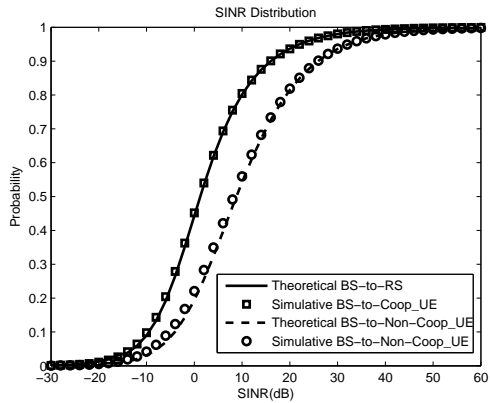}
\caption{CDFs of SINRs for non-cooperative UEs and cooperative UEs.}
\label{fig:SINR}
\end{figure}

Figure \ref{fig:SINR} displays the SINR distributions of cooperative UEs and non-cooperative UEs. The curve of the non-cooperative UE reveals that the simulation result
matches the analytical result well, thus corroborating the accuracy of our theoretical analysis. For the cooperative UE, the theoretical curve denotes the SINR distribution
of BS-to-RS link, which verifies that the SINR distribution of BS-to-UE link can be approximated by that of BS-to-RS link.

\begin{figure} [t]
\centering
\includegraphics[width=2.9in]{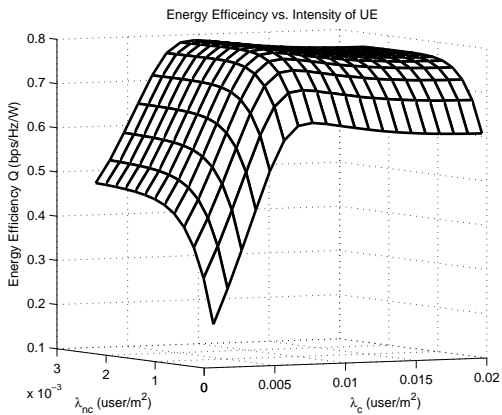}
\caption{Energy efficiency versus the intensities of non-cooperative UEs and cooperative UEs.}
\label{fig:three-dimension}
\end{figure}

Figure \ref{fig:three-dimension} illustrates how the energy efficiency of this network changes with various $\lambda_{nc}$ and $\lambda_{c}$. We observe that the energy efficiency $Q$ increases with $\lambda_{nc}$ and $\lambda_{c}$ until reaching $\!Q_{opt}$ with value 0.78bps/Hz/W and then tends to keep constant after a slight decrease. The reason should be that in the early stage as the intensities of both cooperative UEs and non-cooperative UEs begin to increase, the idle subchannels are sufficient for the existing UEs which results in the increase of the energy efficiency. After reaching $Q_{opt}$, more subchannels become occupied by BSs and RSs which incurs more ``BS'' and ``RS'' tier interference and leads to the decrease of energy efficiency. But soon after the intensity of UEs increases sufficiently large, all the subchannels are persistently occupied by BSs and RSs with the UEs served by time-sharing, and then the interference saturates, which makes the energy efficiency unchanging.

\begin{figure} [t]
\centering
\includegraphics[width=2.9in]{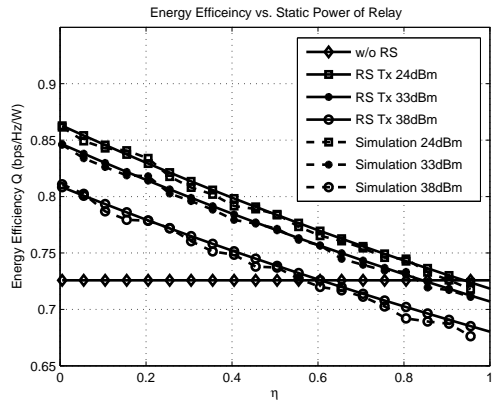}
\caption{Energy efficiency versus the offset power $b_{R}$ and the Tx power of RS.}
\label{fig:static power}
\end{figure}


Figure \ref{fig:static power} shows how the offset power $b_{R}$ and the Tx power of RS affect the energy efficiency. We use the variable $\eta$ to change the value
of $b_{R}$ beginning with the default value 32.0W, i.e., $\!b_{R}\!=\!\eta\!\times\!32.0W\!$. The curves reveal that the simulation results match the analytical results
well which corroborate the accuracy of our theoretical analysis of energy efficiency. The figure shows that the energy efficiency of the RS with 24dBm Tx power case
outperforms that of the 33dBm and 38dBm cases, and the energy efficiency decreases with the increasing of $\eta$. In addition, we get the significant finding here that
there exist thresholds of RS's offset power to compare the energy efficiency metric in the cellular networks with and without relays, which verifies the research
motivation of this paper. Specifically, if the offset power of RS is smaller than the threshold, then it is worthwhile to introduce relays into the cellular networks,
otherwise not. For example, when the Tx power of RS is 24dBm, the thresholds in this case is $\eta_{th}=0.95$, i.e., $b_{R}^{th}=30.4W$, then the relay-assisted networks will
have better energy efficiency when $b_{R}<b_{R}^{th}$ compared with conventional networks without RSs.
\section{Conclusions}
\label{sect:conclusions}
In this paper, we proposed an analytical model for evaluating the energy efficiency of a relay-assisted cellular network using stochastic geometry approach instead of merely relying on system simulations. We introduced coded cooperation as the transmission scheme, and derived the distribution of SINRs and mean achievable rates of both non-cooperative UEs and cooperative UEs. And then the analytical expression of energy efficiency for our network is obtained. In our simulations, a 3D surface of energy efficiency was presented to reveal the relationship among the energy efficiency and the intensity of each type of UEs.
Besides, we observed that the static power consumption of RSs behaves as an important factor to determine whether cooperative relaying can save energy of cellular networks or not. For certain network configurations, there exist static power thresholds of RSs, under which the energy efficiency of the cellular networks can be meliorative.
The work in this paper sheds a useful insight into the real cellular network design and planning.

\bibliographystyle{IEEEtran}
\bibliography{123}

\balance

\label{sect:conclusion}
\end{document}